\documentclass[aps,twocolumn,pra,tightenlines]{revtex4}
\usepackage{amsmath,amsthm,amsfonts,amssymb,amsbsy,mathrsfs}
\usepackage{units,times,float,graphicx,bm,comment,color}
\usepackage[colorlinks,citecolor=blue,linkcolor=blue,urlcolor=blue]{hyperref}
\newcommand{\doi}[2]{\href{https://doi.org/#1}{#2}}
\newcommand{\g}{\textit{g}}

\begin{document}
\title{Nonreciprocal photon blockade induced by parametric amplification in an asymmetrical cavity}
\author{Shao-Xiong Wu\footnote{sxwu@nuc.edu.cn}, Xue-Chen Gao, Huan-Huan Cheng, and Cheng-Hua Bai}
\affiliation{School of Semiconductor and Physics, North University of China, Taiyuan 030051, China}
\date{\today}

\begin{abstract}
We propose a scheme to generate and manipulate nonreciprocal photon blockade effect in an asymmetrical Fabry-P\'{e}rot cavity, which consists of a single two-level atom and a second-order nonlinear medium. By utilizing the intrinsic spatial asymmetry of cavity and applying a parametric amplification pumping laser to the nonlinear medium, we can realize direction-dependent single-photon and two-photon blockade effects. For nonreciprocal single-photon blockade, our proposal is robust across a wide range of parameters, such as the cavity or atomic detuning, coupling strength, and atomic decay. Within similar parameter ranges, nonreciprocal two-photon blockade can be achieved and modulated by finely adjusting the parametric amplification pumping. Our project offers a feasible access to generating high-quality and tunable nonreciprocal single/two-photon source and paves a new avenue for investigating the nonreciprocity of photon quantum statistical properties.
\end{abstract}
\maketitle

\section{Introduction}
Nonreciprocal devices play essential roles in both visible and invisible optical processing, including safeguarding the laser source from the influence of noise and blocking reversed-propagating signals \cite{DLSounas17}. Since typical optical reciprocity is governed by the Lorentz reciprocity theorem, nonreciprocal devices must break the isotropy of optical transmission and statistical properties. Traditional optical isolators based on the Faraday rotation effect tend to be relatively bulky, owing to the drawback of requiring strong magnetic fields. With the rapid advancement of optical information, researchers have noticed that it is worthwhile to explore new ways to break the optical reciprocity. So far, various magnetic-free schemes have been proposed to implement optical nonreciprocal devices, such as systematic spatiotemporal modulation \cite{NAEstep14}, nonlinear optics \cite{YShi15, LDBino18,MCotrufo24,PYang19}, optomechanical interaction \cite{SManipatruni09, GAPeterson17,HXu19,DMalz18,ZShen18}, synthetic gauge field \cite{KFang13,YChen21}, chiral quantum optical systems \cite{MScheucher16,PLodahl17,SZhang18}, utilizing energy loss \cite{XHuang21}, unbalanced coupling \cite{KXia14}, reservoir engineering \cite{AMetelmann15}, and so on.

Previous works on nonreciprocity are primarily focused on the photon classical statistical properties, specifically the first-order correlation function. For the quantum statistical properties, it is expressed in terms of higher-order correlation function. For example, the photons are antibunched when the second-order correlation function is less than 1, a phenomenon known as photon blockade (PB). The PB effect \cite{AImamoglu97} reveals that photons exhibit antibunching phenomenon in their quantum statistical properties, and describes the resonant excitation of the first photon in the system, which prevents subsequent photons from entering and being absorbed. The next photon can only access the system after the current photon has been radiated out. As single-photon sources are fundamental resources with widespread applications in quantum optics and quantum information processing. Methods for generating them, such as utilizing the PB effect, have attracted lots of attention. There are two predominant mechanisms used to interpret the PB effect, namely conventional photon blockade (CPB) \cite{AImamoglu97,KMBirnbaum05} and unconventional photon blockade (UPB) \cite{TCHLiew10,MBamba11}.

The CPB, induced by the anharmonic eigenenergy spectrum and typically requiring a sufficiently strong coupling strength of the system, has been proposed in various systems, including nonlinear cavities \cite{WLeonski94,AMajumdar13}, cavity quantum electrodynamics systems \cite{BDayan08,JTang21,CWang17, YFHan18,RTrivedi19}, and optomechanical systems \cite{JQLiao13,FZou19, DYWang20,JYang21}. It has also been experimentally demonstrated in several platforms \cite{KMBirnbaum05,CLang11,AReinhard12}. The UPB overcomes the limitation of strong nonlinearity by constructing additional excitation paths and employing the destructive quantum interference effect between multiple paths, e.g., introducing optomechanical systems \cite{PRabl11}, auxiliary cavities \cite{HFlayac17}, nonlinearity \cite{YHZhou15}, etc. Furthermore, the optimization of PB, utilizing the synergistic effect of both CPB and UPB mechanisms, has been proposed. Examples include hybrid Kerr nonlinear multimode systems \cite{HZhu23}, quasichiral atom-photon interaction \cite{YWLu22}, and demonstration UPB in superconduct circuit \cite{HJSnijders18,CVaneph18}. In addition to single-photon blockade, two-photon blockade has also been experimentally confirmed \cite{CHamsen17}.

Scholars have started to combine nonreciprocity with PB and explore their quantum properties, i.e., achieving the conversion between bunching and antibunching photons by changing the incident direction, a phenomenon called nonreciprocal PB. Distinct from classical optical nonreciprocity, nonreciprocal PB enables unidirectional control of quantum correlations. Various methods for its realization have been proposed, including spinning systems based on the Fizeau shift \cite{RHuang18,BLi19, YMLiu23,CGou23,HZShen20}, asymmetrical cavities \cite{XXia21,XCGao23}, and cavity optomagnonic systems \cite{HXie22}. Recently, the nonreciprocal PB has been experimentally demonstrated in both the strongly coupled atom-cavity system \cite{PYang23} and the optical microdisk with a coupled scatterer \cite{AGraf22}.

The parametric amplification process can significantly enhance the quantum effects \cite{XYLv15,WQin18,LTang22,SXWu24, HJabri22,DWLiu23,CPShen23,DYWang23,XJWu24,ZHLiu24,WZhang24,HLin24}. In this paper, we will delve into the generation and manipulation of the nonreciprocal PB effect in an asymmetrical Fabry-P\'{e}rot cavity, which contains a single two-level atom and a second-order nonlinear medium. The cavity is driven by a classical coherent laser, while the second-order nonlinear medium is pumped by a parametric amplification laser. Not only nonreciprocal single-photon blockade but also two-photon blockade can be generated under the optimal parametric amplification pumping, which plays central roles in both processes. For the nonreciprocal single-photon blockade, the isolation ratio can be high (such as over 30 dB) across a quite wide range of system parameters, including the detuning of cavity/atom, the coupling strength, and the atomic decay, where are covered by the current experimental techniques. Furthermore, the nonreciprocal two-photon blockade induced by parametric amplification pumping can also be achieved in a similar parameter region by finely adjusting the pumping laser. Compared to previous reports on the nonreciprocal PB effect, our scheme offers the advantage of being stationary and immune to the influence of mechanical motion on this fragile quantum effect. It can flexibly  realize either nonreciprocal single-photon blockade or two-photon blockade as required, thus providing new avenues  for experimentally verifying the quantum statistical properties of photons.

The rest of this paper is organized as follows. In Sec. \ref{sec2}, the physical model and the Hamiltonian of the whole system are introduced; then the dynamics and the optimal pumping condition for the single-photon blockade are derived. In Sec. \ref{sec3}, how to enhance and manipulate the nonreciprocal single-photon blockade effect is discussed in detail. In Sec. \ref{sec4}, the modulation of nonreciprocal two-photon blockade effect is investigated. The discussion and conclusion are summarized in the end.

\section{Theoretical model and dynamics}\label{sec2}
\begin{figure}[t]
\centering
\includegraphics[width=0.8\columnwidth]{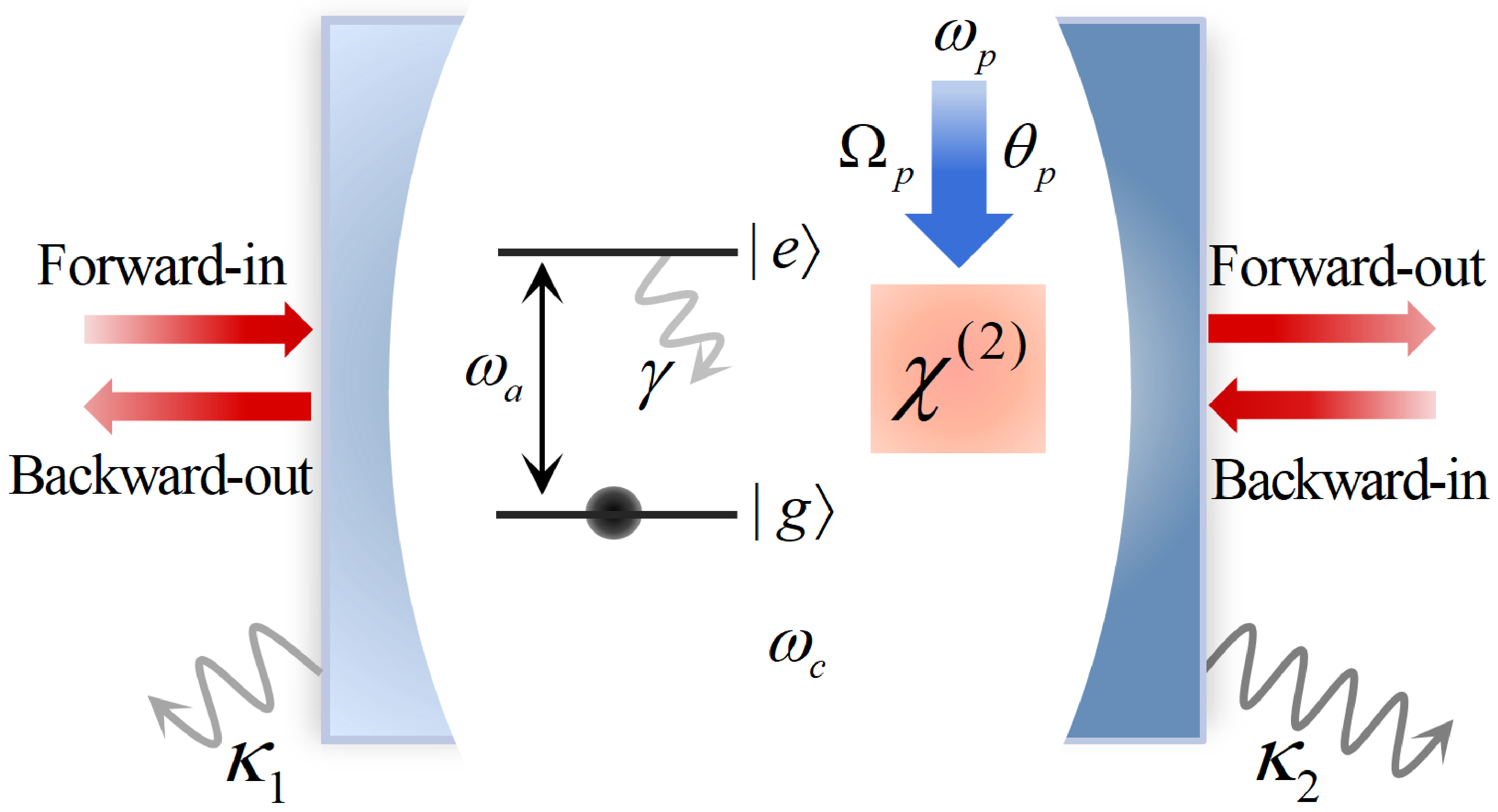}
\caption{Schematic diagram of the asymmetrical cavity-atom system incorporating a $\chi^{(2)}$-type nonlinear medium. $\kappa_1(\kappa_2)$ represents the decay rate of the left (right) cavity mirror. The cavity is driven by a classical coherent laser, while the nonlinear medium is pumped by a parametric amplification laser. The scenario with left input and right output is defined as the forward case, and the reverse scenario is defined as the backward case.}\label{fig:1}
\end{figure}

As depicted in Fig. \ref{fig:1}, the project under consideration comprises a two-level atom and a $\chi^{(2)}$-type nonlinear medium trapped in an asymmetrical Fabry-P\'{e}rot cavity. The asymmetry of the cavity is characterized by the difference between the decay rates $\kappa_1$ and $\kappa_2$ of the left and right mirrors, which are directly related to the reflectivity of the super-mirrors. The total cavity decay rate is expressed as $\kappa=\kappa_1+\kappa_2+\kappa_{\text{loss}}$, where $\kappa_{\text{loss}}$ represents the additional dissipation of cavity mode due to absorption or scattering by the mirrors. Since the cavity is of high quality ($\kappa_{\text{loss}}/\kappa\ll1$), the dissipation $\kappa_{\text{loss}}$ is extremely small and can be neglected, therefore it is reasonable to assume $\kappa=\kappa_1+\kappa_2$ for simplicity. The Hamiltonian of the system can be written as (in units of $\hbar=1$)
\begin{align}
H_s=&\omega_c{a^\dag}a+\omega_a\sigma_+\sigma_-+\g(a\sigma_++{a^\dag}\sigma_-),\label{eq:Hs}
\end{align}
where the first two terms represent the free Hamiltonian of the optical cavity and the atom, and the third term denotes the interaction Hamiltonian between the atom and the cavity with a coupling strength of $\g$, respectively. The operator $a^\dag(a)$ represents the creation (annihilation) operator of the cavity field with resonance frequency $\omega_c$, the operators $\sigma_+=|e\rangle\langle\g|$ and $\sigma_-=|\g\rangle\langle e|$ denote the raising and lowering operators of the two-level atom with an energy transition $\omega_a$ between the excited state $|e\rangle$ and the ground state $|\g\rangle$.

The cavity is driven by a classical coherent laser with frequency $\omega_d$, and the effective driving strength, which depends on the driving direction (marked by $\kappa_i$), is $\Omega=\sqrt{\kappa_i}b_{\text{in}}$ with the driving amplitude $b_{\text{in}}=\sqrt{P_{\text{in}}/(\hbar\omega_d)}$ under the power $P_{\text{in}}$. The $\chi^{(2)}$-type nonlinear medium is pumped by a parametric amplification laser with strength $\Omega_p$, frequency $\omega_p$. The relative phase between the driving laser and the pumping laser is $\theta_p$. The Hamiltonian describing the cavity driving and the parametric amplification pumping is expressed as
\begin{align}
H_d=\frac{\Omega_p}{2}a^{\dag2}e^{-i(\theta_p+\omega_pt)}+\Omega{a^\dag}e^{-i\omega_dt}+\text{H.c.}\label{eq:Hd}
\end{align}
As displayed in Fig. \ref{fig:1}, if the driving laser incidents from the left port of the asymmetrical cavity, it is defined as the forward propagating mode case; conversely, if the driving direction is from the right port, it is defined as the backward case. Due to the intrinsic asymmetry of the cavity, i.e., $\kappa_1$ and $\kappa_2$, the primary distinction between the forward and backward propagating cases lies in the decay rate; therefore we can initially define the forward case, and the backward case can be readily obtained by simple swapping the decay rate $\kappa_1$ with $\kappa_2$.

The total Hamiltonian of the proposed system is given by $H_0=H_s+H_d$. Under the frequency matching condition  $\omega_d=\omega_p/2$, and applying the rotating frame defined by $U=\exp[-i\omega_d({a^\dag}a+\sigma_+\sigma_-)t]$, the transformed total Hamiltonian can be rewritten as
\begin{align}
H=&\Delta_c{a^\dag}a+\Delta_a\sigma_+\sigma_-+\g(a\sigma_++{a^\dag}\sigma_-)\notag\\
&+\Omega(a^\dag+a)+\frac{\Omega_p}{2}(e^{-i\theta_p}a^{\dag2}+e^{i\theta_p}a^2),
\label{eq:HR}
\end{align}
where $\Delta_c=\omega_c-\omega_d$ and $\Delta_a=\omega_a-\omega_d$ correspond to the detunings of the cavity field and the atom with respect to the driving laser, respectively.

Next, we will analyze the statistical properties of photons based on the correlation function, by calculating the quantum state analytically and numerically. On the one hand, the analytical expression of quantum state $|\psi(t)\rangle$ can be derived by solving the non-Hermitian Schr\"{o}dinger equation, where accounts for the influences of dissipations of both the cavity field and the atom. By phenomenologically adding imaginary dissipation terms into Hamiltonian (\ref{eq:HR}), the non-Hermitian Hamiltonian of the system is determined by
\begin{align}
H_\text{non}=&\tilde{\Delta}_c{a^\dag}a+\tilde{\Delta}_a\sigma_+\sigma_-+\g(a\sigma_++a^\dag\sigma_-)\notag\\
&+\Omega(a^\dag+a)+\frac{\Omega_p}{2}(e^{-i\theta_p}a^{\dag2}+e^{i\theta_p}a^2),\label{eq:Heff}
\end{align}
where $\tilde{\Delta}_c=\Delta_c-i\kappa/2$ and $\tilde{\Delta}_a=\Delta_a-i\gamma/2$ denote the complex detunings of the cavity field and the atom, respectively; $\kappa$ is the total decay rate of the cavity, and $\gamma$ is the intrinsic damping rate of the atom.

Under the weak driving condition, i.e., $\Omega_p\ll\Omega\ll\kappa$, the cavity mode will lie in a low-excitation limit, and the dynamical evolution subspace of the system can be truncated to a sufficiently low photon number state, such as $n\leq3$, and the evolved quantum state $|\psi(t)\rangle$ takes the following form
\begin{align}
|\psi(t)\rangle\approx\sum_{n,m=0}^{n=3,m=2}C_{n\g}|n,\g\rangle+C_{me}|m,e\rangle,
\label{eq:psi}
\end{align}
where $C_{n\g}$ and $C_{me}$ are the corresponding probability amplitudes of the states $|n,\g\rangle$ and $|m,e\rangle$, respectively. When the state $|\psi(t)\rangle$ becomes steady, its normalization constant, determined by $\mathcal{N}=\sum_{n=0}^3|C_{n\g}|^2+\sum_{m=0}^2|C_{me}|^2$, is approximated to 1. Substituting Eqs. (\ref{eq:Heff}) and (\ref{eq:psi}) into the non-Hermitian Schr\"{o}dinger equation $i|\dot{\psi}(t)\rangle=H_\text{non}|\psi(t)\rangle$, we can obtain a set of differential equations for the probability amplitudes
\begin{align}
i\dot{C}_{1\g}=&\tilde{\Delta}_cC_{1\g}+\sqrt{2}\Omega C_{2\g}+\g C_{0e}+\Omega,\notag\\
i\dot{C}_{2\g}=&\sqrt{2}\Omega C_{1\g}+2\tilde{\Delta}_cC_{2\g}+\sqrt{2}\g C_{1e}+\frac{\sqrt{2}}{2}\Omega_pe^{-i\theta_p},\notag\\
i\dot{C}_{3\g}=&\frac{\sqrt{6}}{2}\Omega_pe^{-i\theta_p}C_{1\g}+\sqrt{3}\Omega C_{2\g}+3\tilde{\Delta}_cC_{3\g}+\sqrt{3}\g C_{2e},\notag\\
i\dot{C}_{0e}=&\g C_{1\g}+\tilde{\Delta}_aC_{0e}+\Omega C_{1e},\notag\\
i\dot{C}_{1e}=&\sqrt{2}\g C_{2\g}+\Omega C_{0e}+(\tilde{\Delta}_c+\tilde{\Delta}_a)C_{1e},\notag\\
i\dot{C}_{2e}=&\sqrt{3}\g C_{3\g}+\frac{\sqrt{2}}{2}\Omega_pe^{-i\theta_p}C_{0e}+\sqrt{2}\Omega C_{1e}\notag\\
&+(2\tilde{\Delta}_c+\tilde{\Delta}_a)C_{2e}.
\label{eq:gailv}
\end{align}

Under the weak driving limit, one can assume that $C_{0\g}\simeq1$ and ignore the higher-order small quantities, thus the solutions for the probability amplitudes $C_{1\g}$ and $C_{2\g}$ of the steady state are obtained as follows
\begin{align}
C_{1\g}=&\frac{2\Omega(\g^2\tilde{\Delta}_a-\tilde{\Delta}_c\Delta_{ac})+\Omega\Omega_pe^{-i\theta_p}(\g^2+\Delta_{ac})} {2[\g^4+(\tilde{\Delta}_c^2-\Omega^2)\Delta_{ac}-\g^2\Delta_{ac}']},\label{eq:gailvjie}\\
C_{2\g}=&\frac{2\Omega^2(\g^2+\Delta_{ac})+\Omega_pe^{-i\theta_p}[\g^2(\tilde{\Delta}_a+\tilde{\Delta}_c)-\tilde{\Delta}_c\Delta_{ac}]} {2\sqrt{2}[\g^4+(\tilde{\Delta}_c^2-\Omega^2)\Delta_{ac}-\g^2\Delta_{ac}']}.\notag
\end{align}
For simplicity, the variable coefficients $\Delta_{ac}=\tilde{\Delta}_a(\tilde{\Delta}_a+\tilde{\Delta}_c)-\Omega^2$, and $\Delta_{ac}'=\tilde{\Delta}_c(2\tilde{\Delta}_a+\tilde{\Delta}_c)+2\Omega^2$ are introduced in Eq. (\ref{eq:gailvjie}). The other probability amplitudes $C_{0e}$, $C_{1e}$, $C_{3\g}$ and $C_{2e}$ are given in Appendix \ref{appA}.

The phenomenon of PB is associated with the quantum statistical properties of photons. In the case of single-photon blockade, the equal-time second-order correlation function of photons is less than 1, indicting that the photons exhibits antibunching phenomena. According to the quantum input-output theory and by introducing the output field operator $b_{\text{out}}=-i\sqrt{\kappa_2}a$ when the driving laser is injected through the left mirror, the equal-time second-order correlation function of the output photons can be expressed as
\begin{align}
\g^{(2)}(0)=&\frac{\langle b_{\text{out}}^\dag b_{\text{out}}^\dag b_{\text{out}} b_{\text{out}}\rangle}{\langle b_{\text{out}}^\dag b_{\text{out}}\rangle^2}\label{eq:g20}\\
=&\frac{2|C_{2\g}|^2+2|C_2e|^2+6|C_{3\g}|^2}{(|C_{1\g}|^2+|C_{1e}|^2+2|C_{2\g}|^2+2|C_{2e}|^2+3|C_{3\g}|^2)^2}.\notag
\end{align}
For the analytical solution of Eq. (\ref{eq:gailv}), the magnitude of the amplitudes scale as $\{C_{1\g},C_{0e}\}\sim\Omega/\kappa$, $\{C_{2\g},C_{1e}\}\sim\Omega^2/\kappa^2$, and $\{C_{3\g},C_{2e}\}\sim\Omega^3/\kappa^3$. Consequently, Eq. (\ref{eq:g20}) can be approximately expressed as $\g^{(2)}(0)\approx2|C_{2\g}|^2/|C_{1\g}|^4$. Meanwhile, the probability distributions of photons inside the cavity can be approximately represented as $\langle a^{\dag}a\rangle\approx|C_{1\g}|^2$.

To achieve a perfect PB effect, the probability of the system being in the two-photon state should be assumed to be zero, i.e., $|C_{2\g}|^2=0$. According to Eq. (\ref{eq:gailvjie}) and the mathematical inequality $\g^2+\Delta_{ac}\neq0$, there is no valid driving strength $\Omega$ to obtain perfect UPB when the system is off-resonant and the pumping laser is absent. However, when the pumping laser is applied to the nonlinear medium, a perfect PB can occur with the following optimal condition
\begin{align}
\Omega_pe^{-i\theta_p}=\frac{2\Omega^2(\g^2+\Delta_{ac})}{\tilde{\Delta}_c\Delta_{ac}-\g^2(\tilde{\Delta}_a+\tilde{\Delta}_c)}.
\label{eq:optimal}
\end{align}

On the other hand, the time-dependent evolution of the system can be described by numerically solving the Lindblad quantum master equation
\begin{align}
\dot{\rho}=-i[H,\rho]-\frac{\kappa}{2}\mathcal{L}[a]\rho-\frac{\gamma}{2}\mathcal{L}[\sigma_-]\rho,
\label{eq:master}
\end{align}
where $\rho$ denotes the density matrix of the whole system, including both the cavity field and the atom, the form of the Lindblad superoperator for an operator $o$ is governed by $\mathcal{L}[o]\rho=o^{\dag}o\rho-2o\rho o^{\dag}+\rho o^{\dag}o$. The system is assumed to be situated in a zero-temperature vacuum environment.

Due to the optimal condition given by Eq. (\ref{eq:optimal}) and the intrinsic inconsistency in transmission (i.e., decay rate $\kappa_i$) between the super-mirrors 1 and 2, it will lead to different optimal pumping conditions $\Omega_pe^{-i\theta_p}$, which will cause a shift in the PB window and make it possible to control and manipulate the nonreciprocal PB. In the subsequent content, we will investigate the nonreciprocal PB (\ref{eq:g20}) through both the analytical solution of differential equations (\ref{eq:gailv}) and the numerical solution of the master equation (\ref{eq:master}).

\section{Manipulation of nonreciprocal single-photon blockade}\label{sec3}
\begin{figure}[b]
\centering
\includegraphics[width=0.8\columnwidth]{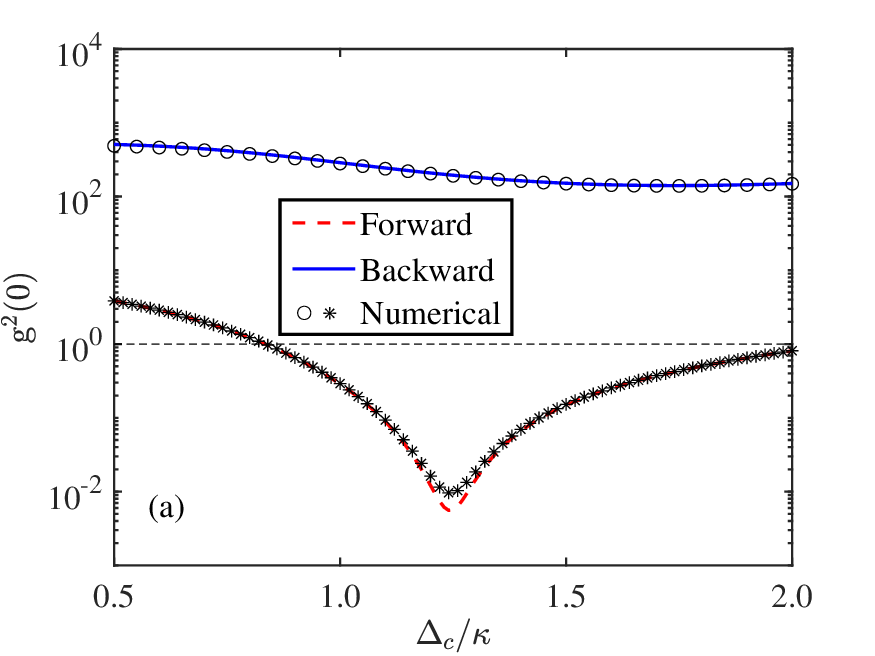}\\
\includegraphics[width=0.8\columnwidth]{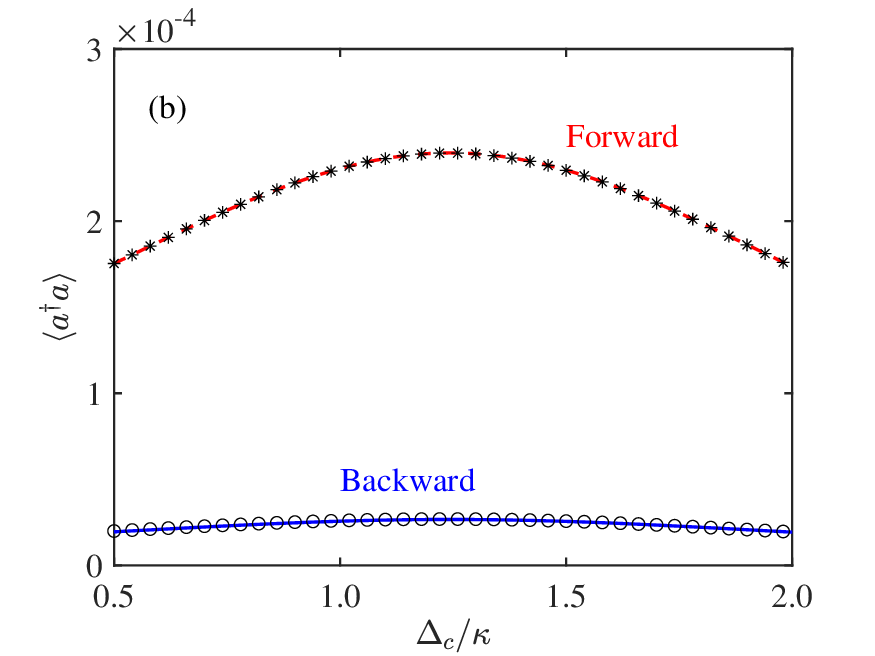}
\caption{(a) The second-order correlation $\g^{(2)}(0)$ and (b) the intracavity photon population $\langle a^{\dag}a\rangle$ versus the cavity detuning $\Delta_c$. The red dashed line is based on the analytical result in Eq. (\ref{eq:g20}) for the forward propagating mode, and the blue solid line represents the backward case. The black asterisk and circle are the corresponding numerical results based on the quantum master equation (\ref{eq:master}). The parametric amplification pumping meets the optimal condition (\ref{eq:optimal}) for the forward case.}\label{fig:2}
\end{figure}

To vividly display the statistical properties of photons in the asymmetrical cavity, we plot the equal-time second-order correlation function $\g^{(2)}(0)$ and the intracavity photon population $\langle a^{\dag}a\rangle$ as a function of the cavity detuning $\Delta_c$ for both the forward and backward propagating modes in Fig. \ref{fig:2}. For the asymmetrical Fabry-P\'{e}rot cavity, the decay rates of the left and right mirrors are set as $\kappa_1/\kappa=0.9$ and $\kappa_2/\kappa=0.1$, respectively; for the atom, the damping rate is chosen as $\gamma/\kappa=0.7$, the coupling strength with the cavity is assumed to be $\g/\kappa=1$, and the atomic detuning is $\Delta_a/\kappa=0.6$. The driving amplitude of the asymmetrical cavity is selected as $b_{\text{in}}/\sqrt{\kappa}=0.02$. It should be noted that the parameters chosen in this project fall within the currently experimentally feasible range. For the D$_2$ line of Cesium, the power of 852 nm continuous driving laser $P_d=\hbar\omega_d\Omega^2/\kappa_i$ is approximately 0.3 fW with cavity decay rate $\kappa/2\pi=3.7$ MHz \cite{PYang23}.

In Fig. \ref{fig:2}(a), the red dashed line represents the second-order correlation function of the forward propagating mode and the blue solid line is the corresponding backward case, which are expressed by the analytical result of Eq. (\ref{eq:g20}); the black asterisk and circle denote the numerical results based on the master equation (\ref{eq:master}). The difference between the forward and backward propagating cases lies solely in the reversal of the driving laser's direction, and the optimal condition (\ref{eq:optimal}) is selected based on the forward case. One can observe that the PB will occur for cavity detuning $\Delta_c/\kappa>0.83$ up to 2 under the forward input, while photons exhibit bunching throughout the entire range for backward incident. Even at the dip of the $\g^{(2)}(0)$ curve, where the perfect UPB is theoretically expected to occur, the value of $\g^{(2)}(0)$ is not zero. The mechanism behind this phenomenon can be explained by the fact that even though the population of the state $|2,\g\rangle$ can be additionally destructive interfered by the pumping transition $|0,\g\rangle\rightarrow|2,\g\rangle$ with strength $\sqrt{2}\Omega_pe^{-i\theta_p}$ under the optimal pumping condition (\ref{eq:optimal}); however, the higher-level states are still inevitably excited with an extremely small probability, such as the state $|2e\rangle$, $|3\g\rangle$, etc. Owing to the substantial difference in population magnitude between the higher-level states and the state $|2\g\rangle$, their influences on the second-order correlation function are typically negligible. However, in the vicinity of perfect PB, i.e. the value of $|C_{2\g}|^2$ is almost zero, the influence cannot be ignored, which makes the analytical expression of $\g^{(2)}(0)$ in Eq. (\ref{eq:g20}) is nonzero even at the point of perfect PB.

The probability amplitude method used to derive the analytical result of Eq. (\ref{eq:g20}) is an approximation of the real physical process, due to the pure state hypothesis and the low photon number truncation in Hilbert space, which leads to strong agreement between the probability amplitude method and the quantum master equation method (\ref{eq:master}), except in the region that characterizes the perfect PB. In general, a steady state after evolution is expected to be a mixed state, so the quantum master equation with higher photon number truncation can provide a more precise description of the properties of photon statistics, particularly in the vicinity of perfect PB. Therefore, we will primarily use the quantum master equation approach to investigate the properties of PB.

The population of cavity photons, denoted as $\langle a^{\dag}a\rangle$, for both forward and backward propagating modes are shown in Fig. \ref{fig:2}(b). One can observe that at the position of perfect PB, the population of cavity photons reaches its maximum under forward incidence. In our model, since the detuning is of the same order of magnitude as both the dissipation and the coupling strength, the effective Hamiltonian $H_{\text{non}}$ (\ref{eq:Heff}) should be used to derive the single-photon resonance condition. The corresponding position of the perfect PB and the maximum photon population can be expressed as $\Delta_c\sim\text{Re}[\g^2/\tilde{\Delta}_a]$, and the single-photon emission efficiency can nearly reach $10^3$ per second. For both the forward and backward cases, the analytical result of the photon population $\langle a^{\dag}a\rangle=|C_{1\g}|^2$ based on Eq. (\ref{eq:gailvjie}) is in excellent agreement with the numerical solution of the quantum master equation (\ref{eq:master}).

\begin{figure}[t]
\includegraphics[width=0.8\columnwidth]{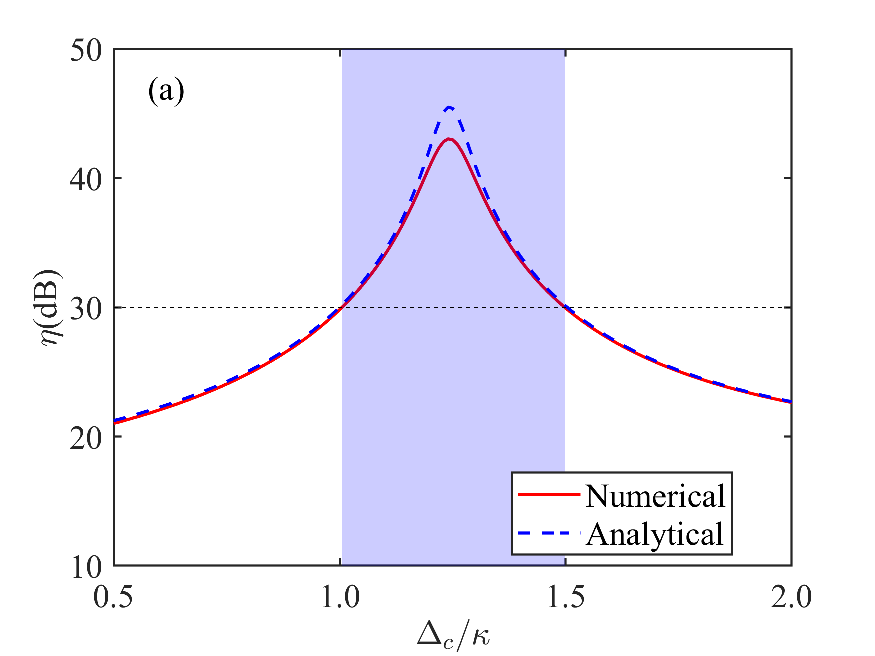}\\
\includegraphics[width=0.8\columnwidth]{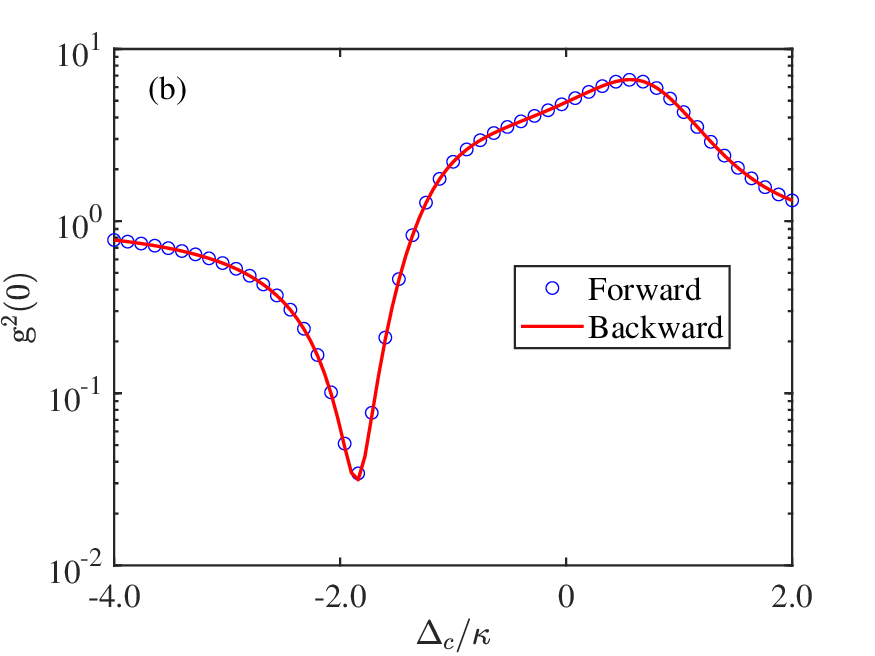}
\caption{(a) Variation of the nonreciprocal ratio $\eta$ with the cavity detuning $\Delta_c$. The blue dashed line represents the analytical result in Eq. (\ref{eq:gailvjie}), while the red solid line depicts the numerical result obtained from the master equation (\ref{eq:master}). (b) The second-order correlation $\g^{(2)}(0)$ for the forward (blue circle) and backward (red line) cases in the absence of pumping laser. The other parameters are the same as those in Fig. \ref{fig:2}.}\label{fig:3}
\end{figure}

With the aim of quantitatively assessing the nonreciprocal PB effect in two counter-propagating modes, the nonreciprocal ratio of PB is defined as:
\begin{align}
\eta=-10\log_{10}\left[\frac{\g_f^{(2)}(0)}{\g_b^{(2)}(0)}\right],
\label{eq:eta}
\end{align}
where $\g_{f/b}^{(2)}(0)$ represents the second-order correlation function for the forward/backward propagating mode. In Fig. \ref{fig:3}(a), we plot the nonreciprocal ratio of PB, where the blue dashed line corresponds to the analytical result (\ref{eq:g20}) and the red solid line is based on the master equation (\ref{eq:master}). Similarly to Fig. \ref{fig:2}(a), the Hilbert space is truncated to the lowest three energy levels in the quantum state (\ref{eq:psi}). As a result, the nonreciprocal ratio calculated using the  probability amplitude method will be higher than that obtained through the master equation method, especially in the vicinity of the perfect PB. However, the range that the nonreciprocal ratio over 30 dB is almost the same, where is indicated by the blue area.

In Fig. \ref{fig:3}(b), the second-order correlation $\g^{(2)}(0)$ is shown based on the master equation (\ref{eq:master}) when the pumping laser is absent. The blue circle represents the forward propagating mode, and the red solid line represents the backward case. It can be observed that although both the two counter-propagating modes can generate PB effect, the positions of perfect PB are almost identical relative to the cavity detuning $\Delta_c$, and the window that PB effect occurred is indistinguishable. So, there is no occurrence of the nonreciprocal PB effect in the absence of the pumping light. When the optimal pumping laser is employed, the nonreciprocal PB can be observed, as shown in Fig. \ref{fig:2}(a). One can conclude that the pumping laser plays the central role in the generation and control of the nonreciprocal PB effect. With the assistance of the optimal pumping laser, one can demonstrate the generation of a high nonreciprocal ratio of PB over a wide range by jointly adjusting the system parameters.

\begin{figure}[b]
\centering
\includegraphics[width=0.8\columnwidth]{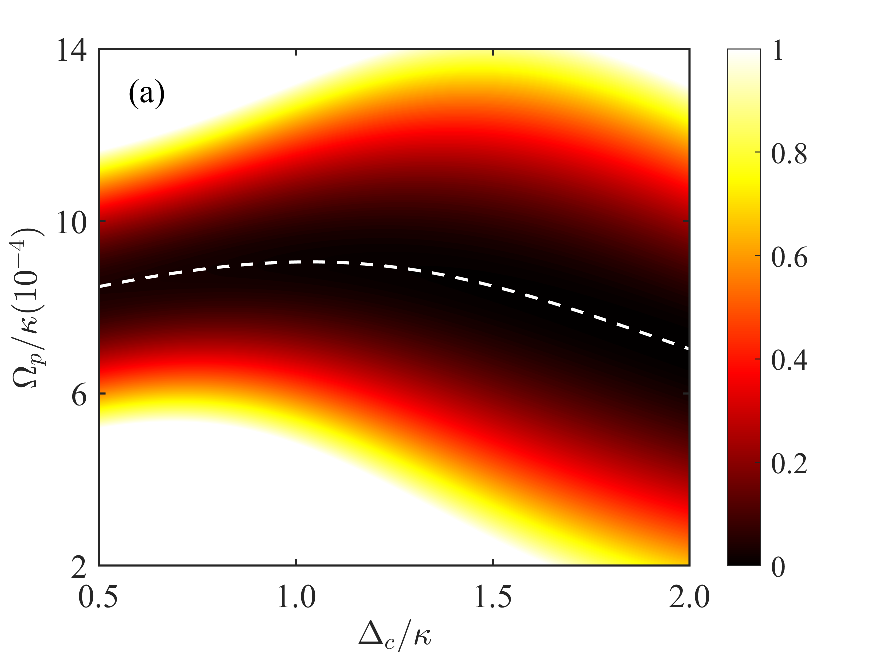}
\includegraphics[width=0.8\columnwidth]{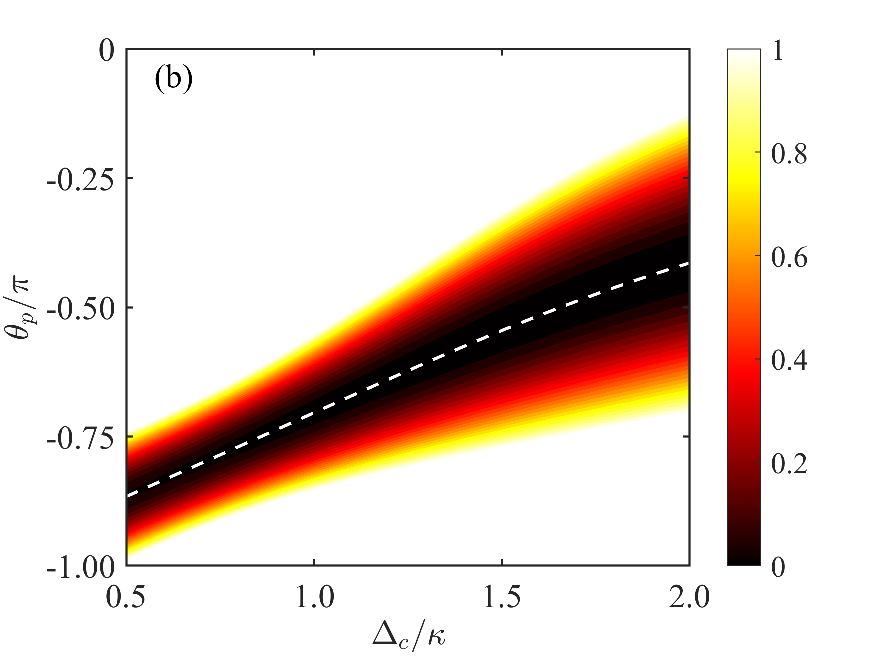}
\caption{The second-order correlation $\g^{(2)}(0)$ versus the cavity detuning $\Delta_c$ and the amplitude $\Omega_p$ (a) and the relative phase $\theta$ (b) of the pumping laser. The dashed white lines indicate the optimal pumping condition (\ref{eq:optimal}). The other parameters are the same as those in Fig. \ref{fig:2}.}\label{fig:4}
\end{figure}

\begin{figure*}
\centering
\includegraphics[width=0.8\columnwidth]{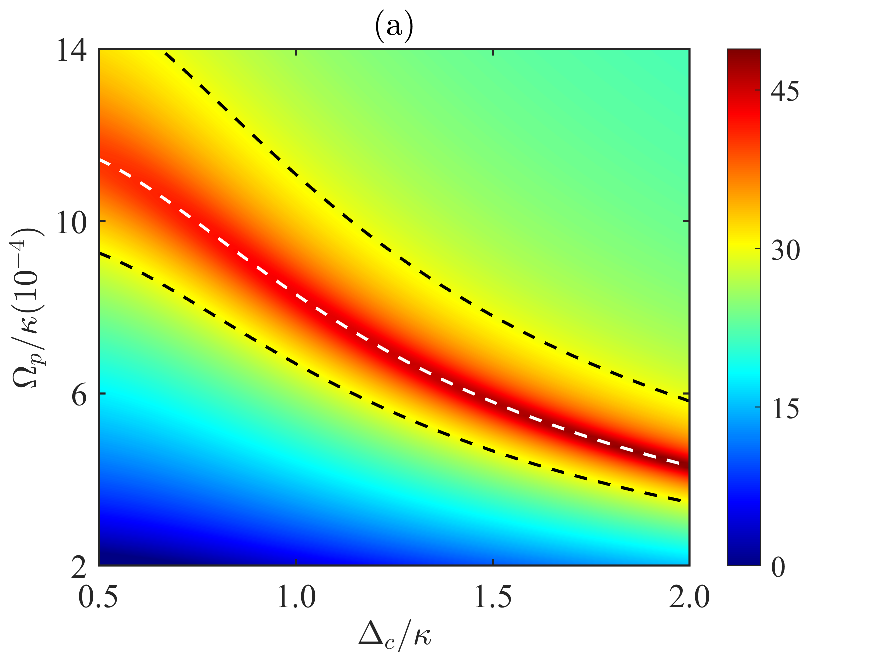}
\includegraphics[width=0.8\columnwidth]{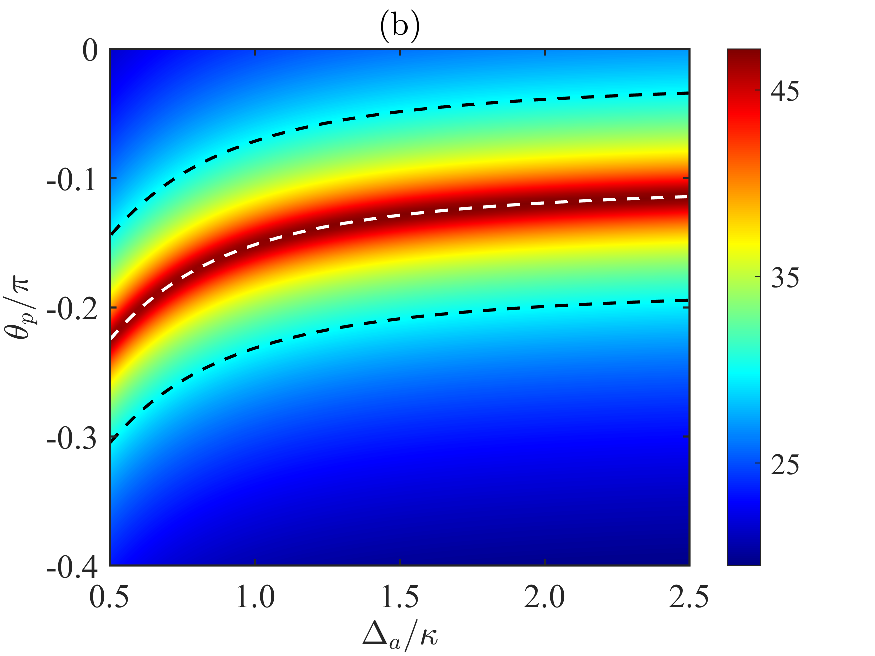}\\
\includegraphics[width=0.8\columnwidth]{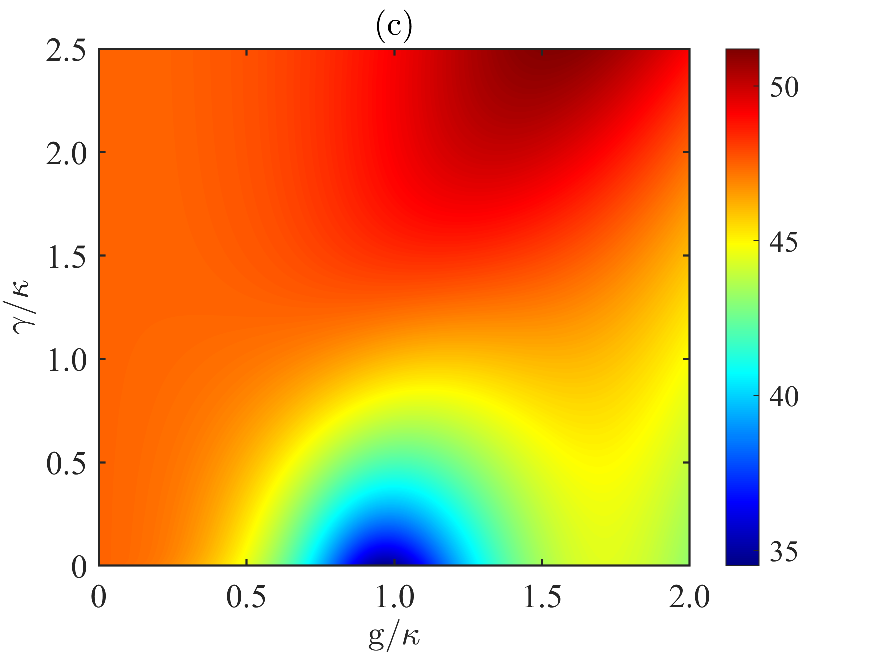}
\includegraphics[width=0.8\columnwidth]{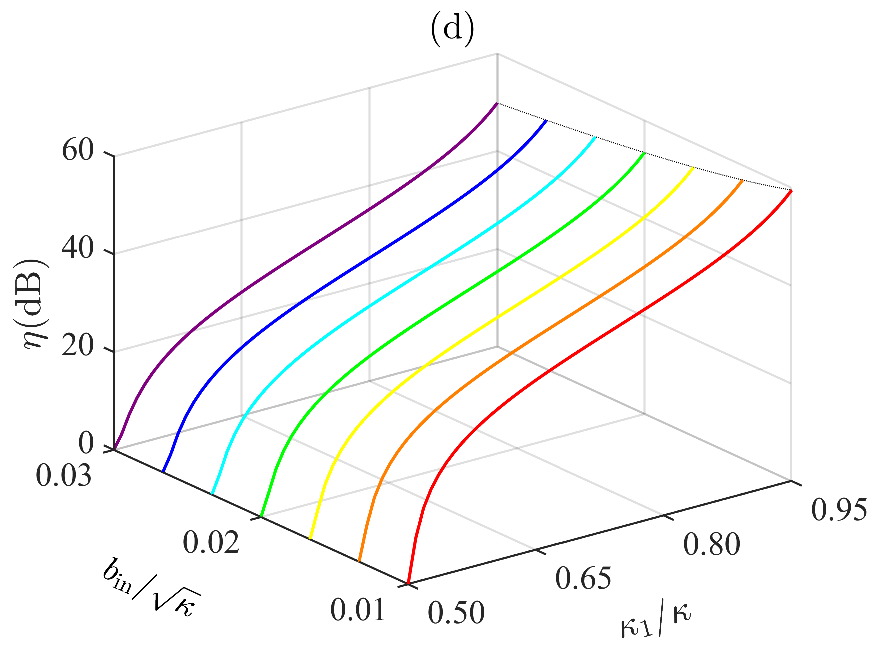}
\caption{(a) The nonreciprocal ratio $\eta$ between the forward and backward cases versus the cavity detuning $\Delta_c$ and the pumping amplitude $\Omega_p$. The coupling strength fixes at $\g/\kappa=0.5$. (b) The nonreciprocal ratio $\eta$ versus the atomic detuning $\Delta_a$ and the relative phase $\theta_p$. The cavity detuning sets at $\Delta_c/\kappa=1.5$. (c) The nonreciprocal ratio $\eta$ versus the coupling strength $\g$ and the atomic decay $\gamma$. (d) The variation in the nonreciprocal ratio $\eta$ along with the cavity decay rate $\kappa_1$ and the driving amplitude $b_{\text{in}}$. The other parameters are the same as those in Fig. \ref{fig:2}.}\label{fig:5}
\end{figure*}

The variation in second-order correlation function $\g^{(2)}(0)$ with the cavity detuning $\Delta_c$, the amplitude $\Omega_p$ and the relative phase $\theta_p$ of the pumping laser is plotted in Fig. \ref{fig:4}. As illustrated in Fig. \ref{fig:4}(a), the range of the pumping amplitude $\Omega_p/\kappa$ is on the order of $10^{-4}$, which is approximately two orders of magnitude lower than that of the driving laser $b_{\text{in}}/\sqrt{\kappa}=0.02$, and is in accordance with the physical principles underlying the weak driving hypothesis in Eq. (\ref{eq:psi}). The white dashed line represents the position of perfect PB theoretically calculated with the optimal pumping condition (\ref{eq:optimal}), which is in good agreement with the numerical result obtained through the master equation (\ref{eq:master}). The optimal relative phase $\theta_p$ is always satisfied as the cavity detuning $\Delta_c$ changes. Similarly, we plot the variation of the second-order correlation function $\g^{(2)}(0)$ with respect to the cavity detuning $\Delta_c$ and the relative phase $\theta_p$ in Fig. \ref{fig:4}(b). The value of optimal relative phase $\theta_p$ is negative, which results from the Hamiltonian (\ref{eq:Hd}) of the pumping laser. The relative phase $\theta_p$ can be modulated by mode-locked lasers in experiment, and the perfect PB varies periodically with the phase $\theta_p$ over a $2\pi$ range. Fig. \ref{fig:4} shows the variation of the second-order correlation function for the forward propagating mode, one can verify that the PB phenomenon does not occur for the backward case under the same system parameters setting. Both the amplitude $\Omega_p$ and the relative phase $\theta_p$ play vital roles in the generation of nonreciprocal PB by utilizing the intrinsic asymmetry of the Fabry-P\'{e}rot cavity.

As shown in the above, the nonreciprocal PB can be observed under appropriate system parameters with the aid of the pumping laser. In Fig. \ref{fig:5}, we illustrate the variation of nonreciprocal ratio $\eta$ of PB with respect to different system parameters. In Fig. \ref{fig:5}(a), we show the nonreciprocal ratio $\eta$ between the forward and backward cases versus the cavity detuning $\Delta_c$ and the amplitude $\Omega_p$ of the pumping laser. The coupling strength between atom and cavity sets at $\g/\kappa=0.5$, and the optimal relative phase $\theta_p$ associated with the amplitude parameter $\Omega_p$ is always satisfied as the cavity detuning $\Delta_c$ changes. The white dashed line corresponds to the position of the maximum nonreciprocal ratio, which coincides with the location where perfect PB occurs. The black dashed line represents the contour line where the nonreciprocal ratio $\eta$ is 30 dB. By finely adjusting the amplitude $\Omega_p$ and the relative phase $\theta_p$ of the pumping laser, the nonreciprocal PB exceeding 30 dB can be realized over a broad range of parameters. In Fig. \ref{fig:5}(b), we plot the nonreciprocal ratio $\eta$ versus the atomic detuning $\Delta_a$ and the relative phase $\theta_p$. The cavity detuning is chosen as $\Delta_c/\kappa=1.5$.

The variation of nonreciprocal ratio $\eta$ in the atomic coupling strength $\g$ and the atomic decay rate $\gamma$ under the optimal pumping condition (\ref{eq:optimal}) is shown in Fig. \ref{fig:5}(c). According to the panel, it appears that a larger atomic decay rate is more favorable for achieving a stronger nonreciprocal PB effect. However, from a perspective of experimental manipulation, a larger atomic decay rate $\gamma$ implies stronger spontaneous emission, which complicates the control and manipulation of quantum state. So, we need to balance the nonreciprocal ratio $\eta$ and the atomic decay rate $\gamma$, and select appropriate parameters. In the vicinity of $\gamma/\kappa\sim1$, it is feasible to achieve a nonreciprocal ratio $\eta$ far exceeding 30 dB. For the coupling strength $\g$, its impact can be basically divided into two parts. In the region where $\gamma/\kappa<1$, a smaller coupling strength ($\g/\kappa<1$) is more favorable for achieving a high nonreciprocal ratio; whereas in the region where $\g/\kappa>1$, a larger coupling strength ($\g/\kappa>1$) is more conducive to attaining a high nonreciprocal PB ratio. Since achieving stronger coupling experimentally often requires more sophisticated and challenging techniques, it is advantageous for experiment realization when the coupling strength is moderation. Based on the current experiment technology, the magnitude order of the atomic decay rate $\gamma$ can be achieved in the same order as the cavity decay $\kappa$, which brings great convenience to experimental verification without the requirement of strong coupling strength.

\begin{figure*}
\centering
\includegraphics[width=0.8\columnwidth]{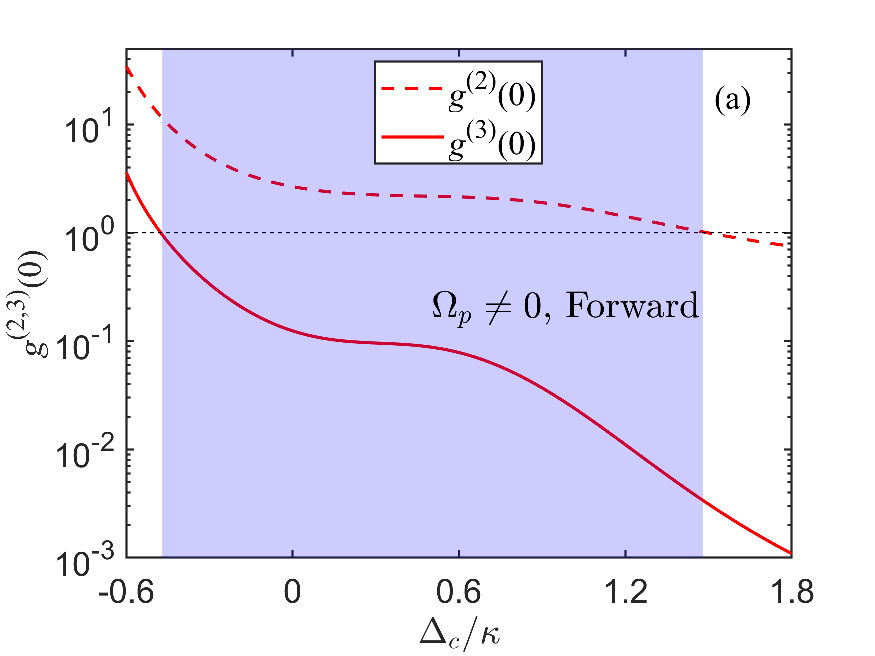}
\includegraphics[width=0.8\columnwidth]{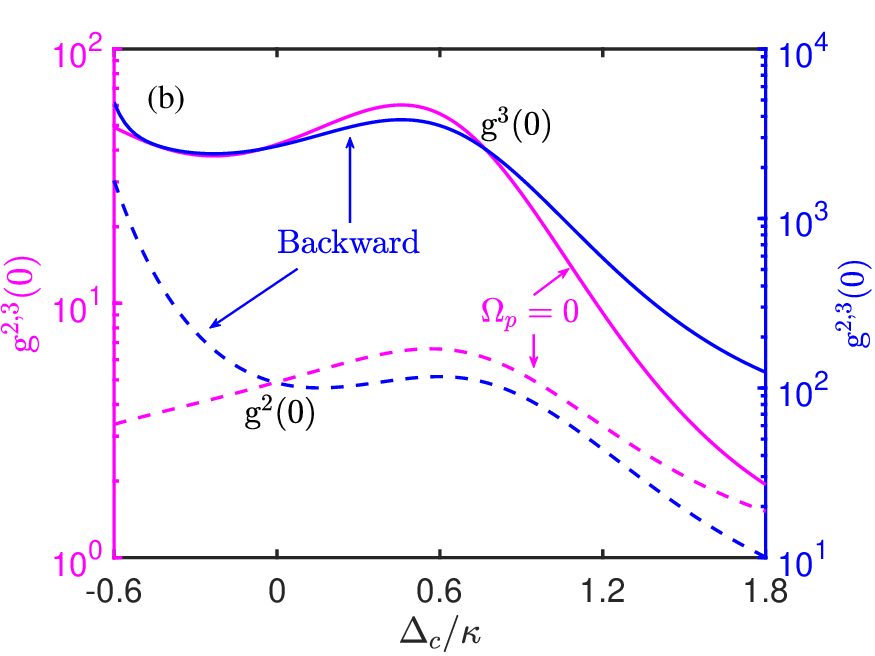}\\
\includegraphics[width=0.8\columnwidth]{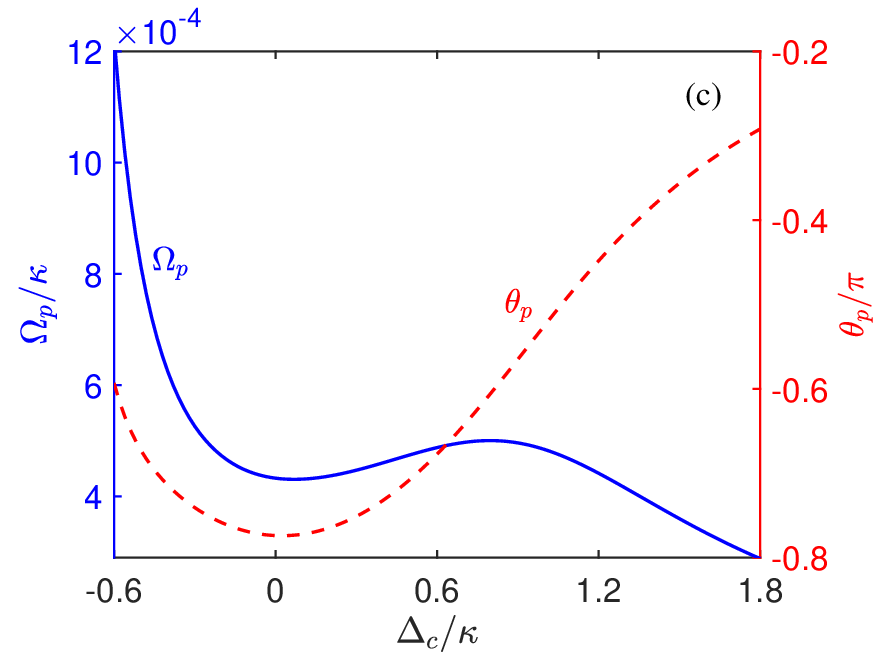}
\includegraphics[width=0.8\columnwidth]{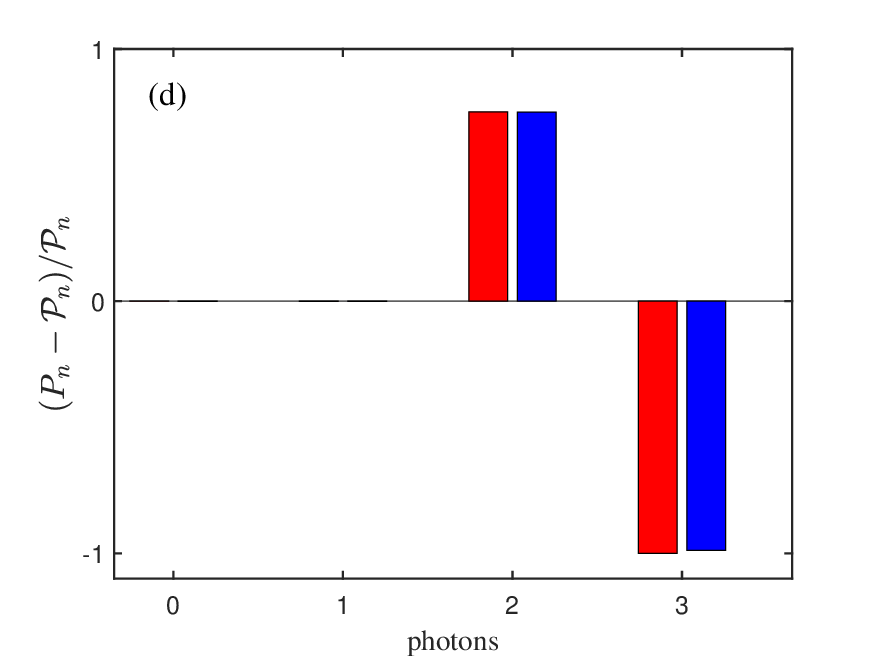}
\caption{(a) The two-photon blockade for the forward case can occur in the light blue shadowed region when the optimal pumping condition (\ref{eq:optg3}) is employed. (b) The second-order correlation function $\g^{(2)}(0)$ and the third-order correlation function $\g^{(3)}(0)$ versus the cavity detuning $\Delta_c$ for the cases of backward propagating mode or the pumping laser is absent. (c) The variation in the amplitude $\Omega_p$ and the relative phase $\theta_p$ of the optimal pumping condition (\ref{eq:optg3}) with respect to the cavity detuning $\Delta_c$. (d) The relative deviation of the photon distribution $P_n$ from the standard Poisson distribution $\mathcal{P}_n$ for the identical mean photon number at $\Delta_c/\kappa=1$.}\label{fig:6}
\end{figure*}

In Fig. \ref{fig:5}(d), the variation in the nonreciprocal ratio $\eta$ is shown along with the cavity decay rate $\kappa_1$ of the left super-mirror and the driving amplitude $b_{\text{in}}$. Due to the asymmetric nature of the Fabry-P\'{e}rot cavity, the rate $\kappa_1/\kappa$ starts from 0.5. We compare the amplitude of driving laser $b_{\text{in}}/\sqrt{\kappa}$ ranging from 0.01 to 0.03. As expect from physical intuition, the nonreciprocal ratio $\eta$ decreases very slowly as the amplitude of the driving laser increases. Since the driving strength $\Omega=\sqrt{\kappa_1}b_{\text{in}}$ is increased, there is a higher probability that the photons in the cavity will be excited into the higher-level states even under the UPB mechanism, resulting in a degradation of the single-photon blockade effect. If the objective is to achieve PB for the forward propagating mode with the assistance of optimal pumping (\ref{eq:optimal}), and to attain the highest possible nonreciprocal ratio $\eta$, it appears that the power of driving laser is preferably minimized and the approximation conditions used in the probability amplitude method (\ref{eq:psi}) will become more effective, causing the value of the second-order correlation function $\g^{(0)}_f(0)$ to decrease and the corresponding maximum nonreciprocal ratio $\eta$ to increase. However, it will also lead to a prominent reduction of the intracavity photon population $\langle a^{\dag}a\rangle$, resulting in a decrease in the efficiency of single-photon emission efficiency. For a concrete example, one can refer to the difference of photon population between the two counter-propagating modes in Fig. \ref{fig:2}(b). Therefore, it is necessary to comprehensively consider both the nonreciprocal ratio and the single-photon generation efficiency, and to set appropriate system parameters accordingly.

\section{Manipulation of nonreciprocal two-photon blockade}\label{sec4}
In our model, we can not only investigate the control of nonreciprocal single-photon blockade assisted by the pumping laser, but also explore the manipulation of nonreciprocal two-photon blockade induced by the parametric amplification process. For two-photon blockade, a suitable criterion requires that the third-order correlation satisfies $\g^{(3)}(0)<1$ and the second-order correlation meets $\g^{(2)}(0)\geq1$. In general, the perfect two-photon blockade can be realized by calculating that $|C_{3\g}|^2=0$, and the optimal pumping condition for the two-photon blockade can be expressed in the form
\begin{align}
\Omega_pe^{-i\theta_p}=\frac{1}{\beta}[\Delta\pm\sqrt{\Delta^2-4\beta\Omega^2(\g^2\zeta+\Delta_{ac}\delta)}],
\label{eq:optg3}
\end{align}
where the coefficients are given by $\Delta=\alpha-4\g^4-\delta_{ac}$, $\delta_{ac}=\g^2[3\tilde{\Delta}_a(\tilde{\Delta}_a+\tilde{\Delta}_c)-4\Omega^2]$, $\alpha=3\tilde{\Delta}_c\Delta_{ac}\delta$, $\beta=2(2\g^2+\Delta_{ac})\delta$, $\delta=\tilde{\Delta}_a+2\tilde{\Delta}_c$, and $\zeta=3\tilde{\Delta}_a+4\tilde{\Delta}_c$, respectively.

For the forward propagating mode, we plot the third-order correlation $\g^{(3)}(0)$ and the second-order correlation $\g^{(2)}(0)$ in Fig. \ref{fig:6}(a) under the assistance of the pumping laser with optimal condition (\ref{eq:optg3}). One can find that the third-order correlation $\g^{(3)}(0)$ (solid red line) is less than 1, while the corresponding second-order correlation $\g^{(2)}(0)$ (dashed red line) is greater than 1 in the blue-shaded region, which is a signature of two-photon blockade. In the same region of cavity detuning $\Delta_c$, we can also easily verify that neither single-photon nor two-photon blockade will occur if the pumping laser is absent (magenta lines) or the direction of the driving laser is reversed (blue lines), as shown in Fig. \ref{fig:6}(b). Comparing Figs. \ref{fig:6}(a) and (b), we can conclude that the manipulation of nonreciprocal two-photon blockade can be realized by resetting the parametric amplification pumping laser carefully.

There is marked difference between the optimal pumping conditions for single-photon blockade (\ref{eq:optimal}) and  two-photon blockade (\ref{eq:optg3}) in terms of both the amplitude $\Omega_p$ and the relative phase $\theta_p$. For the single-photon blockade, the variation in the optimal $\Omega_p$ and $\theta_p$ as a function of cavity detuning $\Delta_c$ is illustrated in Fig. \ref{fig:4} with the white dashed line. The optimal amplitude $\Omega_p$ and the relative phase $\theta_p$ for two-photon blockade (\ref{eq:optg3}) is presented in Fig. \ref{fig:6}(c). In the region of the two-photon blockade window, the pumping amplitude $\Omega_p$ has the same magnitude as that for the single-photon blockade, and the optimal relative phase $\theta_p$ remains negative value with a period of $2\pi$.

In addition to the equal-time $n$-th order correlation function, another widely adopted criterion for evaluating the $n$-photon blockade is based on comparing the relative deviation $(P_n-\mathcal{P}_n)/\mathcal{P}_n$ between the photon number distribution $P_n=\langle{n}\rangle$ and the standard Poisson distribution $\mathcal{P}_n=\langle{n}\rangle^n\exp(-\langle n\rangle)/n!$ with the same average photon number $\langle n\rangle=\langle a^{\dag}a\rangle$. When the photon number distribution and the standard Poisson distribution satisfy $P_n\geq\mathcal{P}_n$ and $P_{n+1}<\mathcal{P}_{n+1}$, the system exhibits $n$-photon blockade effect. To further demonstrate the effectiveness of our project, we plot the relative deviation $(P_n-\mathcal{P}_n)/\mathcal{P}_n$ under the cavity detuning $\Delta_c/\kappa=1$ in Fig. \ref{fig:6}(d), where the red/blue histograms correspond to the analytical/numerical results.  It also indicate that the two-photon blockade phenomenon will occur. It is conveniently to realized the switch between the nonreciprocal single-photon and two-photon blockade using the identical setup by precisely adjusting the amplitude $\Omega_p$ and relative phase $\theta_p$ of the pumping laser.

\section{Discussion}\label{sec5}
In practical parametric amplification process, the efficiency of the parametric down-conversion is inherently limited due to factors such as absorption, scattering, and inherent attenuation in the second-order nonlinear medium. Accounting for these effects is equivalent to introducing an additional dissipation term $\kappa_{\text{loss}}'$ into the Fabry-P\'{e}rot cavity, resulting in the total cavity decay $\kappa=\kappa_1+\kappa_2+\kappa_{\text{loss}}'$, where the additional dissipation term $\kappa_{\text{loss}}'$ cannot be neglected. Nevertheless, the theoretical framework presented in this paper remains fully applicable, with only minor adjustments to the system parameters required.

In a Fabry-P\'{e}rot cavity, the cavity length can  range from a few tenths of a millimeters to several tens of millimeters. The coupling strength $\g=\mu\sqrt{\omega_c/(2\hbar\epsilon_0V_{\text{eff}})}$ between the cavity and the two-level system is determined by the effective mode volume of the Gaussian beam $V_{\text{eff}}=\pi\omega^2L_{\text{eff}}/4$, which depends on both the cavity length $L_{\text{eff}}$ and beam waist $\omega$. Increasing the cavity length $L_{\text{eff}}$ adversely affects the coupling strength $\g$; however, this challenge  may be mitigated by adjusting the radius of curvature of the cavity mirrors to reduce the beam waist $\omega$. By carefully optimizing both the cavity length and mirror curvature, the coupling strength $\g$ can also be maintained within an acceptable range, even when the cavity length is increased. A nonlinear medium such as PPKTP or PPLN, with a customized poling period and quasi-phase matching conditions, can efficiently facilitate second-harmonic generation and optical parametric amplification at specific wavelengths, including 852 nm \cite{JTian16,GHZuo23}. The effective power of the pumping field on the nonlinear medium $P_p^{\text{eff}}=\hbar\omega_p\Omega_p^2/(4\chi)$ is approximately $0.05$ fW, corresponding to a nonlinear single-photon coupling strength of $\chi/2\pi=2.35$ MHz when the efficiency is considered \cite{JLu20}. Furthermore, the relative phase between the driving laser and the pumping laser can be precisely controlled using an electro-optic modulator. Our project is highly feasible and can be implemented using cutting-edge experimental techniques that are widely employed in cavity quantum electrodynamics systems.

In generally, since the photons induced by degenerate optical parametric amplification process in the second-order nonlinear medium is squeezed, the nonreciprocal PB effect can be investigate in a squeezing picture, which is extensively used in the field of quantum optics under the assistance of an additional squeezed vacuum field \cite{XYLv15,WQin18,LTang22}.

\section{Conclusion}\label{sec6}
In conclusion, we have investigated the generation and manipulation of the nonreciprocal photon blockade effect in an asymmetrical Fabry-P\'{e}rot cavity containing a single two-level atom and a second-order nonlinear medium, where the optical parametric amplification is achieved through a pumping laser injected into the nonlinear medium. Due to the intrinsic spatial symmetry breaking of the asymmetrical cavity, the nonreciprocal photon blockade can be generated assisted by the optimal parametric amplification pumping, which serves as the key mechanism in the nonreciprocal process. The influence of system parameters on the nonreciprocal single-photon blockade is comprehensive, and a nonreciprocal ratio exceeding 30 dB can be realized across over a wide range of parameters compatible with the current experiment techniques. Moreover, by precisely adjusting the amplitude and the relative phase of the pumping laser, the nonreciprocal single-photon and two-photon blockade can be switched. Our scheme provides a feasible and practical approach to generate high-quality nonreciprocal single-photon or two-photon sources, which can not only validate nonreciprocal physical mechanisms but also enable potential applications in nonreciprocal photonic devices.

\section*{Acknowledgements}
S.X.W. thanks Dr. Pengfei Yang for insightful  discussions on the experimental feasibility. This work was supported by the National Natural Science Foundation of China (Grant No. 12204440).

\appendix
\section{The probability amplitudes $C_{0e}$, $C_{1e}$, $C_{3\g}$ and $C_{2e}$}\label{appA}
The probability amplitudes $C_{0e}$, $C_{1e}$, $C_{3\g}$ and $C_{2e}$ are also essential for characterizing the quantum properties of photon, which can be given below
\begin{widetext}
\begin{align}
C_{0e}=&\frac{\g\Omega\{2[\tilde{\Delta}_c(\tilde{\Delta}_a+\tilde{\Delta}_c)+\Omega^2-\g^2]-(\tilde{\Delta}_a+2\tilde{\Delta}_c)\Omega_pe^{-i\theta_p}\}} {2[\g^4+(\tilde{\Delta}_c^2-\Omega^2)\Delta_{ac}-\g^2\Delta_{ac}']},\notag\\
C_{1e}=&\frac{g(\tilde{\Delta}_a\tilde{\Delta}_c+\Omega^2-\g^2)\Omega_pe^{-i\theta_p}-2\g\Omega^2(\tilde{\Delta}_a+\tilde{\Delta}_c)} {2[\g^4+(\tilde{\Delta}_c^2-\Omega^2)\Delta_{ac}-\g^2\Delta_{ac}']},\label{Eq:appa}\\
C_{3\g}=&\frac{\Omega(4\g^4\Omega_pe^{-i\theta_p}+c_{31}+c_{32})} {2\sqrt{6}[\g^2-\tilde{\Delta}_c(\tilde{\Delta}_a+2\tilde{\Delta}_c)][\g^4+(\tilde{\Delta}_c^2-\Omega^2)\Delta_{ac}-\g^2\Delta_{ac}']},\notag\\
C_{2e}=&\frac{\g\Omega(c_{21}+c_{22}-c_{23}-c_{24})}{2\sqrt{2}[\g^2-\tilde{\Delta}_c(\tilde{\Delta}_a+2\tilde{\Delta}_c)][\g^4+(\tilde{\Delta}_c^2-\Omega^2)\Delta_{ac}-\g^2\Delta_{ac}']}.\notag
\end{align}
The variable coefficients introduced in Eq. (\ref{Eq:appa}) are $c_{31}=\g^2[2\Omega^2(3\tilde{\Delta}_a+4\tilde{\Delta}_c)+3\tilde{\Delta}_a(\tilde{\Delta}_a+\tilde{\Delta}_c)\Omega_pe^{-i\theta_p} -4\Omega^2\Omega_pe^{-i\theta_p}+2(\tilde{\Delta}_a+2\tilde{\Delta}_c)(\Omega_pe^{-i\theta_p})^2]$, $c_{32}=(\tilde{\Delta}_a+2\tilde{\Delta}_c)[\tilde{\Delta}_a(\tilde{\Delta}_a+\tilde{\Delta}_c)-\Omega^2] [2\Omega^2+\Omega_pe^{-i\theta_p}(\Omega_pe^{-i\theta_p}-3\tilde{\Delta}_c)]$, $c_{21}=2\Omega^4-2(\tilde{\Delta}_a+\tilde{\Delta}_c)(\tilde{\Delta}_a+2\tilde{\Delta}_c)\Omega^2$, $c_{22}=\Omega_pe^{-i\theta_p}\tilde{\Delta}_c[(3\tilde{\Delta}_a+\tilde{\Delta}_c)(\tilde{\Delta}_a+2\tilde{\Delta}_c)+\Omega^2]$, $c_{23}=(\Omega_pe^{-i\theta_p})^2(\tilde{\Delta}_a^2+2\tilde{\Delta}_a\tilde{\Delta}_c+2\tilde{\Delta}_c^2-\Omega^2)$, $c_{24}=g^2[2\Omega^2+\Omega_pe^{-i\theta_p}(3\tilde{\Delta}_c+5\tilde{\Delta}_c+\Omega_pe^{-i\theta_p})]$, respectively.
\end{widetext}

\end{document}